\documentclass[letterpaper,twocolumn]{sig-alternate}
\usepackage{lipsum}% http://ctan.org/pkg/lipsum
\usepackage{graphicx}% http://ctan.org/pkg/graphicx
\usepackage{enumitem}
\usepackage{color}
\usepackage{float}
\usepackage{subfigure}
\usepackage{amsmath}
\pdfoutput=1

\begin{document}

\title{Secure Mobile Identities}

\author{Varun Chandrasekaran \hphantom{xy} Fareeha Amjad \hphantom{xy}Ashlesh Sharma* \hphantom{xy} Lakshminarayanan Subramanian \\
\\
*Entrupy Inc., New York University}
%\authorinfo{\small{
%{Varun Chandrasekaran,} {New York University,}{ varun.chandrasekaran@nyu.edu} \\
%\and
%Fareeha Amjad, {New York University Abu Dhabi,}{ fareeha.amjad@nyu.edu} \\
%\and
%Ashlesh Sharma, {Entrupy Inc.,}{ ashlesh@entrupy.com} \\
%\and
%Lakshminarayanan Subramanian, {New York University,}{ lakshmi@cs.nyu.edu} \\
%}}

\maketitle

\setcounter{secnumdepth}{5}

\begin{abstract} 

The {\em unique identities} of every mobile user (phone number, IMSI) and device (IMEI) are far from secure and are increasingly vulnerable to a variety of network-level threats. The exceedingly high reliance on the {\em weak SIM authentication} layer does not present any notion of end-to-end security for mobile users. We propose the design and implementation of Secure Mobile Identities (SMI), a repetitive key-exchange protocol that uses this weak SIM authentication as a foundation to enable mobile users to establish stronger identity authenticity. 
%The SMI protocol is a spatio-temporal, multi-path, repetitive challenge-response mechanism, enabling two mobile entities to tag every key exchange with a (location, time) pair. It leverages the diversity in this pair to build a reputation for every user-generated key. 
The security guarantees of SMI are directly reliant on the mobility of users and are further enhanced by external trusted entities providing trusted location signatures (e.g. trusted GPS, NFC synchronization points). In this paper, we demonstrate the efficacy of our protocol using an implementation and analysis across standard mobility models. We also demonstrate how the SMI abstraction can enable new forms of secure mobile applications, including messaging, multimedia transfer and a marketplace.

\end{abstract}

\section{Introduction}
\label{intro}

Existing authentication standards adopted by GSM cellular providers provide no notion of end-to-end trust. This is inherently due to the GSM authentication protocol, which is asymmetric in that it does not require the network to identify itself - exposing mobile devices to various network-level security vulnerabilities. Adversaries can launch different forms of Man-in-the-Middle (MitM) attacks \cite{securityVul} including message interception, modification, eavesdropping, spoofing and phishing. The recent proliferation in attacks on GSM security \cite{praattack,germany,comp128} has been exacerbated by the widespread availability of software-defined cellular platforms like OpenBTS \cite{openbts} powered by USRP nodes \cite{usrp}, sysmoBTS \cite{sysmobts}, Fairwaves \cite{fairwaves} and Opencell \cite{opencell}.  Application level security solutions typically rely on simple password based user authentication coupled with SSL/TLS \cite{ssltls}, both of which are fraught with significant problems. Surprisingly, many transactions in the mobile landscape are performed through applications written such that SSL is not always enabled \cite{schrittwieser2012guess}. A recent survey by FireEye on the $1000$ most downloaded free applications on the Google Play Store showed that of the $614$ applications that use SSL/TLS to communicate with a remote server, $448$ ($73\%$) do not check certificates \cite{fireeye}.

We aim to address the following question: {\em Can we build a secure key framework that enables mobile devices to establish an end-to-end trustworthy channel with other mobile devices, or a cloud service provider?} In this paper, we present the design and implementation of {\em Secure Mobile Identities (SMI)}, an identity authentication mechanism that enables mobile devices to convert their weak SIM identity to a stronger {\em self-certifying identity}. Establishing such a secure identity for every mobile device significantly enhances user and application security, especially for cloud-hosted applications that rely on the identity of the mobile device as an alternate trust channel (e.g. 2-factor authentication \cite{2fauth,schneier2005two}). 
%Most 2-factor authentication mechanisms transmit the verification code to the mobile device of a user in clear text providing relatively limited protection.

While there have been numerous efforts to enhance the security of mobile devices at different levels, none of these results focus on building a strong end-to-end trust model around the {\em unique identity} of mobile devices. Most existing solutions have predominantly focused on either the lower level to secure the wireless channel \cite{securepairing, bambos2000channel} or the higher level to protect user identities and applications \cite{lookout, otp, 2fauth}, while others have focused on protecting the mobile device itself \cite{cook2010method}. SMI is built upon the pragmatic assumption that the Mobile Network Operator (MNO) \footnote{Used interchangeably with Cellular Network Provider}is trustworthy, but the underlying cellular channel may not be. The SMI setup models for a local network-level adversary whose aim is to hijack this cellular channel and launch a variety of attacks. Using knowledge of this threat model, the basic building block to establish end-to-end trust between a pair of mobile devices is the {\em probabilistic one-way trust channel} abstraction offered by MNOs. More precisely: {\em At a given time $t$, if a mobile device with a unique identity $D_i$ is genuinely connected to its MNO $O$ and is weakly authenticated (using SIM authentication) by it; and if any arbitrary sender sends a control message $M$ addressed to $D_i$ which successfully reaches $O$'s network; then $M$ is correctly delivered to the device $D_i$ with high probability.} The delivery (and consequently security) guarantees are probabilistic in that successful delivery is dependent on the premise that the channel to $D_i$ is not subject to MitM threats at time $t$.

The SMI protocol involves repetitive key-exchanges using this probabilistic one-way trust channel. Devices can exchange their keys using a challenge-response protocol from a varied set of locations at different times, providing {\em spatio-temporal diversity}. SMI enables a mobile device (or cloud) to build sufficient trust in the key of another device using a reputation mechanism; higher the user mobility, larger the diversity and hence greater the trust on the corresponding (identity, key) pair. Additionally, we describe how a variant of SMI can quickly authenticate identities using third-party identity verification services for a pair of mobile devices, each of whom consider the other a stranger. 

In short, SMI is a mechanism for augmenting the existing SIM security in a cheap and convenient manner. Security guarantees of SMI can be
significantly enhanced through the use of {\em trusted} locations. With minimal infrastructural change, SMI can obtain such trusted location information from a variety of sources. Specific portions of the protocol can be implemented using alternative paths \footnote{Used interchangeably with channel} which include WiFi access points, data or other proximate physical synchronization points, increasing protocol scalability. We demonstrate the robustness of SMI based on proof of concept testing of secure applications on top of the SMI substrate and detailed analysis of its security properties. In summary, SMI provides a predominantly decentralized protocol to build a strong and secure mobile identity with strong end-to-end trust guarantees in the face of network-level security threats.

\section{Setup and Threat Model}
\label{probsettings}

In this section, we define the problem we set out to solve, establish our assumptions and threat model. We begin by briefly reviewing current GSM authentication and how the SMI protocol augments it.

\subsection{Motivation}
\label{threats}

\noindent{\em Background:} Subscribers of any MNO initially procure a Subscriber Identity Module (SIM) which has a unique International Mobile Subscriber Identity (IMSI) and a shared key required for authentication. The Authentication Center (AuC) of the MNO gains knowledge of this key during SIM registration. The SIM and the AuC use an unidirectional challenge-response protocol to verify the shared key, and establishes a temporary session key for the subscriber.

\noindent{\em Threat Vectors:} However, this asymmetric cellular authentication layer is vulnerable to message interception, phishing, eavesdropping and spoofing, among other threats. One such threat is the use of fake base transceiver stations (fBTSs) to trick the device into connecting to an adversarial (henceforth termed fake) network. Such attacks are fairly easy to execute with the availability of software defined radio platforms that can emulate cellular base (transceiver) station functionality \cite{opencell, openbts, openbsc, usrp, sysmobts, fairwaves}. One possible attack involves the fBTSs exhibiting greater signal strength than the real base station to lure mobile subscribers during their first connection to the cellular network. Another possible attack is to selectively jam specific frequency bands in a local area to force mobile devices to search for alternative base stations, some of which could be fake. 

Forcing a mobile device to connect to a fake network exposes it to a myriad of network-level threats. Meyer et al. \cite{germany} show a MitM attack on Universal Mobile Telecommunications Systems (UMTS) by exploiting the lack of integrity protection in the base stations. Kostrzewa et al. \cite{thesis} exploit weaknesses of the A5/2 \cite{comp128} cipher to demonstrate another MitM attack. The A3 and A8 algorithms used by cellular networks, specifically COMP128 and COMP128-1, are known to have flawed implementations which make it easier for network adversaries to launch MitM attacks against \cite{toorani2008solutions}. Another attack is for the fBTS to use known ciphertext attacks \cite{kca} to recover the session keys without disrupting the original subscriber authentication process \cite{barkan2003instant}. Given physical access to a SIM card, there have been attacks that can recover the shared key embedded in the SIM through effective challenge-response mechanisms. Extensions of these attacks have facilitated over-the-air (OTA) cracking of shared keys \cite{nohl,mathur2008radio}. 
%Refer to Appendix \ref{netsec} for more threats.

\subsection{Problem Definition}
\label{definition}

SMI aims to establish an end-to-end secure channel between a pair of mobile devices in a decentralized manner. The problem can also be modified to build trust between a mobile device and a cloud service provider. Both problems require building upon the weak authentication mechanism provided by cellular networks. To elaborate further, consider mobile devices $U_1$ and $U_2$ with unique identities $D_1$ and $D_2$. $U_1$ and $U_2$ generate their own self-certifying \footnote{Complete certification is achieved as trust grows} public keys $PK_1$ and $PK_2$ and their corresponding private keys. Our goal is to achieve secure key exchange between $U_1$ and $U_2$ in the face of network adversaries who may try to disrupt this process. To state succinctly, $U_1$ has to learn the identity-key mapping $(U_2, PK_2)$ and likewise for $U_2$. With SMI, we aim to achieve two properties:

\textbf{Robustness of Key Establishment.} A combination of user and device IDs creates a certified identity as a consequence of repetitive spatio-temporally diverse interactions. This identity requires substantial (computational and economic) expenditure to forge.

\textbf{Scalable Key Exchange.} The protocol scales with the number of participating users. 

\subsection{Adversary Model}
\label{admodel}

Any user can be in the presence of a local network adversary at a particular location. The adversary can create an fBTS to monitor and alter any or all communication relayed through it \cite{stхhlberg2000radio}. With its devices, the adversary can overhear, intercept, and inject any wireless communication. The adversary can also jam specific signal frequencies. We assume that there is reasonable bootstrapping cost associated with setting up the equipment. Hence, we assume that the adversaries are not omnipresent with respect to the locations the user moves to. For practical considerations, we assume that the adversary will neither follow the user to more than $k$ locations, nor will follow any mobile user continuously. We also assume that the adversary is computationally bounded and would require reasonable amount of time (at least few hours \cite{dunkelman2010practical} to few days) to cryptographically break the weak authentication layer of GSM. Therefore, a new mobile subscriber who may immediately be subject to an attack by an adversary has sufficient time to establish and partially prove its self-certifying identity, unless it is subject to denial of service.

\section{Overview}
\label{smprot}

In this section, we present an overview of the SMI protocol. We begin by discussing the assumptions required to support our protocol, followed by explanation of the various components that constitute it. We base our discussion on the case where pairs of mobile devices authenticate their identities with respect to each other, but similar arguments can be made for a mobile device authenticating its identity to the cloud.
 
\subsection{Assumptions}
\label{powt}

The SMS channel operates over the same control channel used for authenticating mobile subscribers. Given the weak authentication mechanism currently in place, the SMS channel can be viewed as a {\em probabilistic one-way trust (POWT) channel} (as defined in \S \ref{intro}). This pragmatic assumption is critical to successful execution of the protocol. We also make the following assumptions, which naturally lead to the construction of the POWT channel assumption. First, we explicitly assume that the cellular service provider and mobile device is trusted. The SMI protocol primarily deals with network-level adversaries who aim to seize the cellular channel to launch MitM attacks. Second, we assume that the shared key present in the SIM issued by the cellular provider is not compromised at the time of issue. If adversaries have the capability to issue their own pre-authenticated SIM cards, then the POWT channel assumption does not hold. Third, we assume that SMS is delivered with high probability as the network layer inside a cellular network is not tapped or tampered internally, barring the last-hop where it is exposed. 

\subsection{Protocol Summary}
\label{basic}

The SMI protocol is a repetitive key-exchange that utilizes user mobility across different locations and time to incrementally establish trust over the POWT (SMS) channel. For the adversary to achieve success, it (in the form of an fBTS) has to be in the vicinity of the mobile device (or cloud provider) at all times. Assisted by this locality constraint placed on the adversary, successful spatio-temporally diverse key-exchanges stipulates an exponential decrease in the ability of an adversary to spoof an (identity, key) pair. Each mobile device participating in the SMI protocol computes a reputation score for every mobile device that it exchanges keys with, where the reputation grows with increased location diversity and number of exchanges. Similarly, the trusted cloud provider builds a reputation of participating devices based on the combination of their unique mobile footprint and device identities using the key-exchange protocol. The security guarantees of the SMI protocol strengthen with increase in the reputation score. 

\vspace{1mm}
\noindent{\bf Mobile To Mobile:} To explain in detail, one user termed the {\em initiator} calculates the reputation score of a participant across discretized time intervals called {\em epochs}. Successful key-exchanges, initiated both periodically and aperiodically across an epoch comprise of trading a signed digest (for verifying message integrity) and an encrypted tuple (ensuring information privacy). The tuple consists of spatial and temporal information coupled with user and device identities. The reputation score is increased when there is adequate diversity in location information and time. When the reputation score reaches a predefined threshold, the identity of the participant is authenticated with respect to the initiator. 

\vspace{1mm}
\noindent{\bf Cloud To Mobile:} This variant is an extension of the previous case. The cloud provider both periodically and aperiodically probes the mobile device for the information tuples, across multiple epochs. Succesful probing results in the growth of the reputation of the device in the eyes of the provider. The key differences with the earlier discussed variant stem from the fact that the cloud provider is trusted, and does not require to prove the mapping between its identity and the keys it uses. It is also important to note that providers are often static (or immobile), in safeguarded locations.

\vspace{1mm}
To maintain a healthy reputation score across each epoch, messages specially tagged for the SMI protocol are repeatedly exchanged in the background over the SMS channel. Running in the background reduces the amount of user intervention thereby increasing the usability of the system. 

\subsection{Spatial Diversity: Location Information}
\label{location}

Mobility ensures that the required geo-diversity is obtained for successful reputation growth. Mobility also enables a user to potentially move outside the range of an adversary. The other component required for spatial diversity, {\em location}, has two major components, namely:

\textbf{Untrusted Locations:} 
%As discussed earlier in \S \ref{smprot}, location plays a pivotal role in facilitating growth of reputation. 
This can be obtained from various sources such as location sensors on mobile devices, GPS, cell towers or other external sources (such as an access point) but these locations, by definition are non-certifying and are easy to forge \cite{tippenhauer2011requirements}. 
%This is further explained in detail in Appendix \ref{usage}.

\textbf{Trusted Locations:} Location information obtained from the sources listed above can be modified or forged. The security, and convergence time of the SMI protocol can substantially improve using time-certifying trusted locations. This certification by an accredited party increases credibility of the location and hence contributes greater towards the score calculation. The different ways of obtaining trusted location information are:
\vspace{-2mm}
\begin{itemize}
\itemsep-0.23em 
\item {\em Trusted GPS Sensors:} Phones could easily be embedded with trusted GPS sensors with pre-certifying keys that can provide time-certifying locations \cite{saroiu2010sensor}.

\item {\em Trusted Identity Verifier:} Recent proliferation of mobile payment systems such as Android Pay or Apple Pay can be used to provide a signed time, location contract. 

\item {\em Trusted BTS:} We envision that by imposing (minimal) hardware changes using trusted co-processors \cite{yee1995secure}, location information from BTS' can be used to verify the location, even during handoff.
\end{itemize}
\vspace{-2mm}

\subsection{Temporal Diversity: Epochs}
\label{time}

An epoch is a discrete, fixed-length time interval, usually in the order of hours. We select this specific duration of time as any adversary will require at least few hours to break the weak GSM authentication. We define an \textit{interaction} as user $U_1$ sending a message to user $U_2$ (or provider $P$), within an epoch. The duration of each interaction is finite and upper bounded. Thus, the number of interactions that can occur in an epoch is also finite. An interaction is \textit{successful} if it hasn't been interfered with by the adversary, through denial of channel or interception. This measurable nature of an interaction allows for revocation mechanisms in case of failure. In an epoch, the initiator begins interactions with other participants in both periodic and aperiodic manners. This aperiodicity or \textit{random probing} prevents an adversary from using knowledge of protocol periodicity for malicious intents. The reputation score is built by successful interactions across multiple epochs. If a user is inactive for more than half an epoch, successful interactions in that particular epoch do not contribute towards the calculation of the reputation score. Thus, an epoch can either complete successfully or abort, with no transient state in between.

\subsection{Reputation Score}
\label{score}

An important concept required for this score computation is that of a {\em zone}, which is defined to be a geographic region with a fixed boundary. A fundamental assumption is movement between zones entails much higher mobility than movement within a zone. The reputation score can be computed as a linear combination of scores from various components, including: (i) weights associated with both trusted and untrusted locations, (ii) third-party identity verifiers providing proofs of identities, (iii) temporal diversity in key-exchanges, and (iv) other meta-factors including node priority (defined in \S \ref{eval}). We believe that this formulation (discussed in detail in \S \ref{reputation}) necessitates user mobility which reduces the likelihood of an attack, and thereby increases the credibility of a user's identity. A threshold is chosen based on the user's mobility model, and frequency of adversarial interference. In essence, this threshold should ensure that the reputation of a user grows at a favorable rate independent of moderate adversarial interference, such that reaching the threshold implies that the user has reasonably mitigated threats from an adversary.
\section{Key-Exchange Protocol}
\label{scp}

In this section, we describe the key steps in the SMI key exchange protocol across epochs where pairs of mobile users sign spatio-temporally diverse tuples of information using self-generated, self-certifying key-pairs.

\vspace{1mm}
{\bf Three-Way Handshake:} The basic building block for SMI is a simple variant of the standard challenge-response mechanism used to establish a connection between two participating entities.  The initiator ($U_1$) and participant ($U_2$) with corresponding device identities $D_1$ and $D_2$ generate key-pairs ($PK_{1},QK_{1}$) and $(PK_{2},QK_{2})$ respectively. $U_1$ and $U_2$ exchange their public keys $PK_1$ and $PK_2$ with a challenge-nonce $c_1$ and a response-nonce $c_2$ and corresponding signatures $S_1(.)$ and $S_2(.)$. Each subsequent key-exchange is associated with a (location, time) pair. These subsequent three-way handshakes can be represented as:
\vspace{-2mm}
\begin{align*}
\textbf{1}.&& \quad U_1 \rightarrow U_2: & f_{PK_2}(L_{1},t_1,U_1,D_1,c_1) \\
\textbf{2}.&& \quad U_2 \rightarrow U_1: & S_2(c_1), f_{PK_1}(L_{2},t_{2},U_2,D_2,c_2,c_1) \\
\textbf{3}.&& \quad U_1 \rightarrow U_2: & S_1(c_2),f_{PK_2}(c_2)
\end{align*}

where $f_{PK_1}$ and $f_{PK_2}$ represent standard public key encryption function, and $S_1(.)$ is a standard signature. Additionally, $(L_1,t_1)$ and $(L_2,t_2)$ represent location time pairs. Issues raised by the rare event of message delivery failure can be resolved using standard retransmission techniques; delivery failure after certain number of retransmissions results in abort for the epoch.

\subsection{Using Untrusted Locations}
\label{dp}

In an epoch, a pair of devices {\em consistently} interact; at the start of an epoch, the users create and exchange seeds which inter-link subsequent interactions within the epoch. We label these seeds to be exchanged as {\em random parameters} and refer to this stage of the protocol as the {\em dialing phase}. This is followed by a {\em connection phase} where there is a repetitive exchange of keys with varied (location, time) pairs.

To elaborate, each epoch begins with users $U_1,U_2$ generating random parameters $a,b \in N$ respectively. These parameters assist in ensuring consistent interactions, defined to be a challenge-response mechanism with the Markov property i.e. the response at any time
$t_{m}$ depends only on the interaction at time $t_{m-1}$. Additionally, for any particular epoch, all further interactions between the users contain {\em incremental signatures} where every signature in an interaction $m$ builds upon the signature in interaction $m-1$. Random parameters $a$ and $b$ are repetitively used as challenge-response nonces for each such signature. %A tuple of information consists of a combination of location information, time information and user, device identities. 
After the dialing phase, both users are familiar with both $a$ and $b$. This is followed by $k$ two-way key exchanges, where at the $2i^{th}$ iteration, $U_1$ sends a signature $S_{2i}(a||S_{2i-1}(.), L_{2i},t_{2i})$ and $U_2$ responds with $S_{2i+1}(b||S_{2i}(.), L_{2i+1},t_{2i+1})$, where $||$ denotes concatenation. 
%Presence of the signature from the previous interaction enforces connectivity within an epoch. %In this context, $a||b||..$ denotes an concatenation based on the number of the iteration (i.e. $2i+1$ for $S_{2i+1}$). For example, if the number of the interaction is $3$, $a||b||..$ is $a||b||a$. The maximum length of the string $a||b||..$ is fixed to be some constant $l_{max}$ and this is maintained using a sliding window of the same size. The window begins to slide once the string length exceeds $l_{max}$, with each successful interaction. 
Hence all the signatures in an epoch are connected and an adversary who has to disrupt has to participate in an entire epoch, right from its inception. The exact key exchange steps in an epoch are outlined below. It is important to note that signatures are indexed by the interaction number. All odd numbered signatures in the dialing phase are generated by user $U_1$. However, in the connection phase, odd numbered signatures are generated by user $U_2$.
%\textbf{} & \quad U_2 \rightarrow U_1: f_{PK_1}(L_{2},t_{2},U_2,D_2,b,S_{init})\\
\vspace{-1mm}
\begin{align*}
\textbf{} & \quad ~\textbf{Dialing Phase:} \\
\textbf{1}.& \quad U_1 \rightarrow U_2: f_{PK_2}(L_{1},t_{1},U_1,D_1,a) \\
\textbf{2}.& \quad U_2 \rightarrow U_1: S_2(a),f_{PK_1}(L_2,t_2,U_2,D_2,b,a)) \\
\textbf{3}.& \quad U_1 \rightarrow U_2: S_3(b||a),f_{PK_2}(b) \\
\textbf{} & \quad ~\textbf{Connection Phase:} \\
\textbf{} & \quad ~\text{$For$ $steps$ $4,5$; $\forall$ $i: i = 2$ to $k+1$} \\
\textbf{4.1} & \quad U_1: S_{2i}(.)= S_{2i}(a||S_{2i-1}(.), L_{2i}, t_{2i}) \\
\textbf{4.2} & \quad U_1 \rightarrow U_2: S_{2i}(.),f_{PK_2}(L_{2i},t_{2i},U_1,D_1,S_{2i}) \\
\textbf{5.1} & \quad U_2: S_{2i+1}(.) =S_{2i+1}(b||S_{2i}(.), L_{2i+1}, t_{2i+1}) \\ 
\textbf{5.2} & \quad U_2 \rightarrow U_1: S_{2i+1}(.),f_{PK_1}(L_{2i+1},t_{2i+1},U_2,D_2,S_{2i+1})\\ 
\textbf{6}.& \quad ~\text{If all $k$ exchanges are successful} \\ 
\textbf{} & \quad U_1: IncreaseRep(D_2, PK_2)\\
\textbf{} & \quad U_2: IncreaseRep(D_1, PK_1)\\
\end{align*}
\vspace{-6mm}

%The dialing phase is a minor modification of the earlier explained three-way handshake to include the random parameters to be exchanged. 
During the connection phase, each interaction between $U_1$ and $U_2$ is initiated both periodically and aperiodically. The former is done to ensure devices have sufficient mobility; the latter to ensure randomness in an epoch exchange, reducing any advantage an adversary could have gained using the knowledge of periodicity. The adversary can forge the identity of the initiator and randomly start another epoch, taking undue advantage of this aperiodicity. We ensure that this is avoided by sending a special tag at the end of the ongoing epoch with consistent interactions. The inception of a new epoch will be approved by the participant only on receiving this tag. An adversary attempting to decrypt the secure and consistent interaction between the users at a time $t_i$ ($i \textgreater 0$) would have to possess knowledge of the random parameters which effectively serves as an increment counter. The probability that the adversary can correctly guess both parameters is very low ($\approx 1/|2^N|^2$) where $N$ is the number of bits in the key space, making it practically impossible to disrupt the epoch except forcing the mobile devices to abort the epoch. Spoofed messages sent by the adversary are \textit{inconsistent} with the ongoing interactions and this alerts the users of its presence in the communication channel. Upon successful completion, both parties enhance their reputation of the individual identities ($IncreaseRep(.)$) while an abort does not have a negative outcome.

\subsection{Using Trusted Locations} 
\label{tl}

The previous version of the protocol was specifically for untrusted locations. In \S \ref{location}, we discussed different types of third-party providers or physical interactions that can provide a notion of trusted locations with verifiable time-bound signatures. Consider one such provider $X$ who has an independent PKI rooted infrastructure \cite{herzberg2000access}. $X$ has a wide distribution of trusted device end-points where each device is embedded with (a) key-pair generated by the root $X$, and (b) corresponding trust certificate from the certificate root of $X$. In addition, each device has a trusted GPS sensor that provides the location $L$ of that particular device. Thus, this {\em donor} device is
equipped to provide a trusted signature $s(L,t)$ to any interacting device. This signature also encompasses the trust certificate of the donor from the root $X$. If one trusts the public key of the root of $X$, then the location signature $s(L,t)$ is a verifiable proof. Except the case of trusted GPS sensor within each mobile device, protocol participating devices obtain these trusted location through physical proximity-based interactions with the donor device. In SMI, we impose that the integrity of this signature persists over the time to live (TTL) window $\Delta_i$ from the moment it was exchanged (say $t_i$) i.e. exchange $i+1$ should be initiated at $t_{i+1}$ such that $t_{i+1} \leq t_i + \Delta_i$. Most mobile devices which connect to the genuine cellular network receive time updates from the network; hence an assumption about minimal time synchronization errors across participating devices and trusted location endpoints is pragmatic. Borrowing terminology from \S \ref{scp}, consider the following example of a three-way handshake (dialing phase) between user $U_1$ and trusted third-party $T$, where $\Rightarrow$ denotes communication using a close-proximity (eg. NFC) channel: 

%The basis of the trusted location key-exchange is that any location information $L_i$ (which is a part of the $i^{th}$ exchange) is associated with a verifiable proof of that location, $s(L_i, t_i)$. Except the case of trusted GPS sensor within each mobile device, protocol participating devices obtain these trusted location through physical proximity-based interactions with the donor device. Any trusted location signature is associated with a limited time value. To successfully utilize this signature in the reputation calculation, the next exchange should be within the TTL window of the signature i.e. exchange $i+1$ should be initiated at $t_{i+1}$ such that $t_{i+1} \leq t_i + \Delta_i$. Borrowing terminology from \S \ref{scp}, consider the following example of a three-way handshake (dialing phase) between user $U_1$ and trusted third-party $T$, where $\Rightarrow$ denotes communication using a close-proximity (eg. NFC) channel:
\vspace{-2mm}
\begin{align*}
\textbf{1}.& \quad T \Rightarrow U: f_{PK_U}(s(L_{T},t_{T}),U_T,D_T,c_1) \\
\textbf{2}.& \quad U \rightarrow T: S_U(c_1),f_{PK_T}(L_U,t_U,U_U,D_U,c_2,c_1)) \\
\textbf{3.1}.& \quad T \rightarrow U: S_T(c_2) \\
\textbf{3.2}.& \quad T \Rightarrow U: f_{PK_U}(c_2) \\
\end{align*}
\vspace{-6mm}

An alternative variant ensures that before providing a trusted location signature $s(L_T,t_T)$ in step $1$, the external location provider can perform an independent three-way handshake with the mobile device in physical proximity to it, where the external trusted location synchronization point uses the SMS channel to verify the authenticity of the device identity $D_U$. In other words, the trusted location provider can use an independent network interface to send the challenge over the SMS channel instead of Bluetooth or NFC channel. Thus, the trusted location provider verifies the interacting device identity $D_U$ subject to the POWT Channel assumption. An even stronger variant would be for the location provider to provide $s(D_T, L_T, t_i)$ where it provides a device specific time-bound location signature. 

\subsection{Using Multiple Channels}
\label{multi-path}

Another simple, but extremely useful variant of the key-exchange protocol is to leverage multiple communication channels during an epoch: SMS, data channel, WiFi, Bluetooth etc. Though the SMS channel is integral for the POWT Channel assumption to hold, it is severely inhibited by its 160 character limit. To partially alleviate load on the SMS channel, we can use alternative channels to offload non-critical data and use it to only send the challenge-response information. One cannot completely bypass the use of the SMS channel for challenge or response messages since that is the only channel that is directly connected to the identity of the mobile device participating in the protocol. This variant of the protocol can help in significantly reducing the number of SMS messages that we need to exchange to establish trust with other devices. 

\subsection{Validation With The Cloud} 
\label{cv}

This version of the protocol specifically discussed the issue of bootstrapping trust with a cloud service provider $P$. The key technical details, especially the random probing component, remain the same; minor variations arise due to the following reasons: (a) The cloud service provider has a fixed location i.e. $L_P$ never changes (as in \S \ref{tl}). This location is often not disclosed to the public to ensure secrecy, ergo providing better protection, (b)  The cloud service provider also has access to a cellular channel with much larger bandwidth than conventional cellular devices, partially mitigating any DoS attack attempts, and (c) The cloud service provider is trusted. It does not have to prove the authenticity of its (identity, key) mapping. It is pragmatic to assume that the cloud is equipped to quickly deal with any infrastructural failures, and proactively defend threats against it. Borrowing previously used terminology, the connection phase of the protocol can be summarized as follows: 

\vspace{-2mm}
\begin{align*}
\textbf{1}.& \quad P \longmapsto U: f_{PK_U}(.) = f_{PK_U}(s(L_{P},t_{P}),U_P,D_P) \\
\textbf{2}.& \quad U \rightarrow P: S_U(f_{PK_U}(.)),f_{PK_P}(L_U,t_U,U_U,D_U)) \\
\textbf{3}.& \quad P \longmapsto U: f_{PK_U}(.) = f_{PK_U}(s(L_{P},t_{P}),U_P,D_P) \\
\textbf{4}.& \quad U \rightarrow P: S_U(f_{PK_U}(.)),f_{PK_P}(L_U,t_U,U_U,D_U)) \\
\end{align*}
\vspace{-6mm}

where $\longmapsto$ denotes a random probe by the cloud server. This variant is particularly useful, when the cloud can host an identity verifier (explained in \S \ref{tpiv}). It is important to note that across all the variants discussed, the variable used to denote time, $t$, varies with each step. 

\section{Reputation}
\label{reputation}

In this section, we further discuss the various components associated with the reputation score. We also discuss obtaining the threshold above which the identity is considered to be reputed, the security guarantees we provide and a quick mechanism to bootstrap reputation using a trusted third-party identity verifier. 

\subsection{Reputation Score Formulation}
\label{formulation}

To begin formalizing the reputation score in detail, let the duration of epoch $i$ be represented by $[t^i_{0},t^i_1]$ where $t^i_{0}$ is the time at inception of epoch $i$ and $t^i_1$ is time at its conclusion. The score for user $U$ at time $t \in [t^i_{0},t^i_1]$ is given by:

\vspace{1.5mm}
\framebox{\large {\centerline{\textbf{$R_U(t,i)$ = $\gamma.f_1(t,i)$ + $\delta.f_2(t)$}}}}

\vspace{1mm}
\begin{itemize}[leftmargin=-.01in]
\itemsep-0.23em 
\item $f_1(t,i)$=$\alpha.(m(t,i))$+$(1-\alpha) .f_1(t^{i-1}_1,i-1)$; smoothing factor $\alpha \in (0.5,1)$ and $i \ge 1$ such that

$m(t,i)$=$c_1.m_1(t,i)$+$c_2.m_2(t,i)$; $c_2 \textgreater c_1 \textgreater 1$, where (a) $m_1(t,i)$ denotes the number of \textit{successful} interactions in the time interval $[t^i_{0},t]$ due to untrusted locations, and (b) $m_2(t,i)$ denotes the number of \textit{successful} interactions with \textit{unique} trusted locations in the same interval. 

\item $f_2(t)$ is a cumulative score associated with the total number of proofs of identity from third-party identity verifiers in the time period $[t^1_{0},t]$ (explained in \S \ref{tpiv}). 

\end{itemize}
\vspace{-2mm}

We set $\alpha \in (0.5,1)$ to provide greater weightage to current transactions compared to historical transactions. We also set $c_2\textgreater c_1$ to provide greater credibility for trusted locations over untrusted locations. The notion of {\em unique} trusted locations provides verifiable proof of user mobility between these locations. Additionally, we view the reputation contribution due to third-party identity verifiers to be larger and hence, the contribution of function $f_2$ outweighs that of $f_1$. This is enforced by constants $\delta \textgreater \gamma$. 
The reputation score after $q$ epochs, $R=X_U + Y_U$ where:

\vspace{2mm}

{\large \centerline{\textbf{$X_U = \alpha.\gamma.\sum_{i=1}^q [(1-\alpha)^{q-i}].m(t^i_1,i)$}}}

\vspace{2mm}

{\large \centerline{\textbf{$Y_U = \delta.f_2(t^q_1)$}}}

\vspace{3mm}
%$X_U$ represents the total reputation gathered across epochs using untrusted locations and trusted locations excluding trusted third-party identity verifiers which is captured by $Y_U$.

\subsection{Management \& Scheduling}
\label{manager}

The number of messages that a mobile device can send is restricted because of cost factors and the network load. Each device leverages the multiple channels available at its disposal to reduce message overhead by transmitting non-critical information using alternative channels which are less constrained (in terms of content size) than the SMS channel. Every device has an inbuilt reputation manager that tracks the reputation of all the user identities in the address book. Additionally, SMI measures the frequency of interaction between users to determine the importance level (priority) of users in the contact list. Users can also pre-specify importance level of other users in their contact list. 

{\em Scheduling:} Based on the importance of a user, their reputation score, and threshold, SMI uses a simple optimization mechanism to determine an {\em active list} of recipients with which SMI will enter into a new epoch of key-exchanges. Given a limited message quota for a time period (characterized by costs), the schedule optimization mechanism orders messages based on the highest utility metric of user priority i.e. important users are to be bootstrapped faster. Once an initiator commences an epoch, it makes sure that it gives precedence for all message exchanges within the epoch even if it may slightly over-run the specified message quota. The reputation manager performs the important function of scheduling to achieve the dual goals of reducing message costs and maintaining the reputation levels of important mobile users in the address book. Apart from minimizing cost, this optimization provides an additional benefit of reducing cellular network load whilst producing the same guarantees with respect to security and score growth. 

\subsection{Security Guarantees}
\label{threshold}

An adversary is bound by the following constraints in the SMI protocol: First, any adversary that aims to hijack an identity has to generate its own key-pair for the identity. To be able to leverage the forged key-pair, the network adversary has to be in continuous proximity of the mobile device and completely prevent any communication between the device and the cellular network. This kind of adversarial behavior is severe and assumed to be infrequent. Second, if the adversary aims to hijack an identity before the device participates in the SMI protocol, then the device can generate a conflicting self-certifying key (discussed in \S \ref{extensions}) which forces the cellular network to abandon/reject the SIM used by that device. Third, an adversary aiming to hijack any epoch must obtain the random parameters exchanged at the beginning of the epoch, an event that occurs with very low probability. Outside of these options to hijack the channel, the easier strategy for the adversary is to jam specific exchanges in an epoch, enough to facilitate epoch abortion. This may at best, delay the establishment of a sufficient reputation score between an initiator, participant pair. All these factors clearly indicate the difficulties an adversary faces in disrupting the SMI protocol over every epoch. 

\vspace{1mm}
Consider a scenario where a network adversary disrupts a particular channel with a small probability $p$. Assuming no collusion between various adversaries, the probability that the user's communication is disrupted across $k$ locations (under the coverage of distinct cell towers) is $p^k$. The adversary controls a fraction of the the cell towers (locations) that the user frequents. We assume that even in the extreme case where the user frequents all cellular towers (of size $n$) within a region, the adversary can control upto half these towers i.e $k \leq n/2$. We also assume that the adversarial nodes are operating in the {\em always on} mode, to immediately detect and hinder any user activity. The value of $p^k << 1$ for larger values of $k$. Thus, with increased mobility, we are able to ensure that the adversarial interference is minimal with high probability (assuming the adversary does not actively follow the user). 

\vspace{1mm}
However, human mobility patterns can be predicted with high precision, and are often repetetive. The adversary can utilize this historical knowledge to strategically position his fBTS nodes to hinder user activity most effectively. For example, the adversary may utilize knowledge that the user is at work for nearly a quarter of his day (nearly $6$ hours), and place a node in the vicinity. As explained earlier, the adversary can combine a DoS attack and identity forgery to build reputation on behalf of the user. However, since the adversary is (economically) constrained to a few locations, the user is notified of its presence once he is not within adversarial proximity. To aleviate the degradation caused by the adversary, the user is presented two options: (a) Slowly build trust through alternate channels till he is out of the vicinity of the adversary, and (b) Visit trusted third-party verifiers (refer \S \ref{tpiv}) to quickly bootstrap trust. 

\vspace{1mm}
With each successful interaction in this period, the reputation score of the participating users grow, building confidence in their identities. To account for reputation decay and to maintain liveliness, nodes need to periodically perform key-exchanges to preserve their reputation above the threshold.

\subsection{Third-Party Identity Verifiers}
\label{tpiv}

A {\em third-party identity verifier} represents a specific variant of the SMI protocol where one of the end-points participating in the key-exchange protocol is not a mobile device but a cloud-based service (such as Keybase, Onename) that maintains reputation gained by a mobile device. A third-party identity verifier can be thought of as a simple datastore that maintains a history of all prior interactions of a device. Any external cloud-based entity can function as a third-party verifier and one can imagine an ecosystem with multiple such verifiers. As this third-party is trusted, a participating device can quickly build its reputation utilizing (historical) information contained. During key exchange, the verifier randomly probes the honest user at different time periods to ensure that the user does not behave erroneously under the influence of a network adversary. 

We envision the following trusted third-party providers to be used in the quick bootstrap process: (a) Cellular Providers like AT\&T, Sprint and Verizon; (b) Independent third-parties like Verisign, Keybase and Onename; (c) Trusted wallet providers like Google, Apple and Mastercard. There are several important use-cases for such third-party verifiers. First, each cellular network can run their own instance of the verifier. This enables the cellular network
to quickly learn the self-certifying key of its mobile users. With sufficient information, the cellular network has essentially established a decentralized PKI for all its mobile users. Consequently, this enables the cellular network to quickly detect SIM cards that may be subject to attacks by network-level adversaries as they present conflicting information. Second, third-party verifiers who are popular and potentially trustworthy can help a pair of mobile users to quickly bootstrap trust between themselves. Finally, trusted wallet providers can use the reputation score to create a reliable mapping between wallet identities, user identities and secure mobile identities of devices of users. 

%\subsubsection{Quick Reputation Bootstrap}
%\label{frb}
%
%Consider the case where two devices $U_1$ and $U_2$, as independent initiators, use the SMI protocol with a trusted third-party provider (eg. Keybase, Onename). After successful key-exchanges, both $U_1$ \& $U_2$ build a score, say $R_1$ and $R_2$, relative to this third-party
%and have authenticated their IDs to it. However, these devices have minimal reputation with respect to each other. 
%
%To quickly bootstrap their reputation of the other, two mutually distrusting devices can perform a three-way handshake to first exchange their keys and obtain a proof of interaction with a trusted-third-party (eg. Keybase, Onename) along with their established reputations certifying by the third-party with supporting proofs of historical key exchange interactions. Using this information, the two devices can quickly bootstrap their reputation of each other as a function of the reputation scores and how much the devices trust the third-party verifier. This quick bootstrap function can significantly reduce the number of message exchanges required to bootstrap and maintain reputation between a pair of devices.

\section{Reputation Score Evaluation}
\label{eval}

In this section, we discuss the various experiments performed, our results for the same and conclude with the key takeaways from our experiments. 
%\end{itemize}

\subsection{Emulation Setup}
\label{setup}

We evaluate the SMI protocol across different mobility models, testing with traces generated by standard mobility models \cite{smooth,citymob}, namely:

\vspace{1mm}
\noindent{\em Simple Traffic:} Models vertical and horizontal mobility patterns without direction changes.

\vspace{1mm}
\noindent{\em Random Walk:} Models unpredictable movement. It is also referred to as Brownian Motion \cite{hida1980brownian}.

\vspace{1mm}
\noindent{\em Probabilistic Random Walk:} This model introduces user-defined probabilistic direction changes.

\vspace{1mm}
\noindent{\em Manhattan Mobility Model:} Nodes move in horizontal or vertical direction on an urban map. It employs a probabilistic approach in the selection of movements. 

\vspace{1mm}
\noindent{\em Downtown Mobility Model:} Adds traffic density to the Manhattan model. Mobility is reduced in predefined downtown areas, where the user is 70\% of the time.

\vspace{1mm}
\noindent{\bf Test Ecosystem:} Our setup comprises of several initiators and participants, totalling $10^4$ nodes. The emulation is run across a $10^{10}$ sq. meter (3861 sq. mile) grid, roughly the size of Los Angeles County, California. In the context of our experimental setup, we define zones to be $10^8$ sq. meter (38.61 sq. mile) grids, totaling $10^2$ different zones. In this setup, each of the $10^2$ zones have a unique distribution of trusted locations. The mean speed chosen for mobility was 14 mph. This was the weighted average value of the speed whilst moving in a vehicle (25 mph) and whilst walking (3 mph), both amidst moderate traffic density. The algorithm chosen as part of the reputation manager is fixed-priority preemptive scheduling \cite{audsley1995fixed}. Any interaction can abort based on adversarial interception of the channel at the location the initiator/participant is at. We set an epoch to be $1$ day in all our experiments, with messages exchanged at an hourly frequency. Therefore, $3$ epochs or nearly $70$ exchanges increases the probability of successful protocol completion to $(1-p)^{70}$. We observe that the average reputation score increase per epoch is $1680$. This produces a corresponding threshold of $5000$ ($1680 \times 3$). Some of the other parameters required to compute the reputation score are explained below:

\vspace{1mm}
{\em Priority:} A manually entered value associated with each participant which indicates its relative importance on the initiator's contact list. Priority ranges from 1 which is the smallest value to 10 which is the largest. As per the scheduling algorithm chosen; in the vicinity of any trusted location, the reputation score of the highest priority nodes is increased.

\vspace{1mm}
{\em Batch:} A uniformly assigned value to each participant creating disjoint participant sets. This ensures that each disjoint set is entitled to score growth uniformly across the entire time interval independent of its priority, thereby preventing stunted score growth. The total number of batches is a function of the number of contacts on the initiator's device.
%\end{itemize}
\vspace{1mm}

\subsection{Composite Mobility Model}
\label{composite}

Due to the volume of data required for evaluation and the difficulty in obtaining anonymized mobility traces, we emulate our experiments using a composite mobility model. This model postulates that the mobility model of each node may change to one of two {\em foreign} models (Manhattan \& Downtown Manhattan) at every six hour interval during a day, but return to its {\em home} mobility model (Simple Traffic) at the start of each day i.e. between 00:00 hours and 06:00 hours, where the node is assumed to be stationary (or mobile in a restricted space). This composite mobility model draws insights from \cite{isaacman2012human,becker2013human,song2010limits,musolesi2007designing} and we believe that this draws some parallels with realistic mobility patterns as it attempts to emulate habitual human mobility.

\subsection{Results}
\label{results}

%In this section, we discuss our results for the various experiments performed. 

\begin{table}[H]
\centering
\begin{tabular}{| l | l | l | l |}
\hline 
\textbf{Mobility} & \textbf{Low} & \textbf{Nil}& \textbf{High}\\ 
\textbf{Model} & \textbf{(p=0.28)} & \textbf{(p=0)}& \textbf{(p=0.54)}\\ 
\hline

 Simple Traffic & 191 & 150.6 & 229.2\\ 
 \hline
 Prob. Random Walk & 175 & 140 & 212.6\\ 
 \hline
 Random Walk & 100.2 & 79.1 & 122.6\\ 
 \hline
 Manhattan Traffic & 142 & 111.9 & 172.5\\ 
 \hline
 Downtown Manhattan & 119 & 92.5 & 141.3\\ 
 \hline
\end{tabular}
\caption {Average Number Of Exchanges Required}
\label{table:avg}
\end{table}

\begin{figure}[!h]
  \begin{center}
      \includegraphics[width=.80\columnwidth]{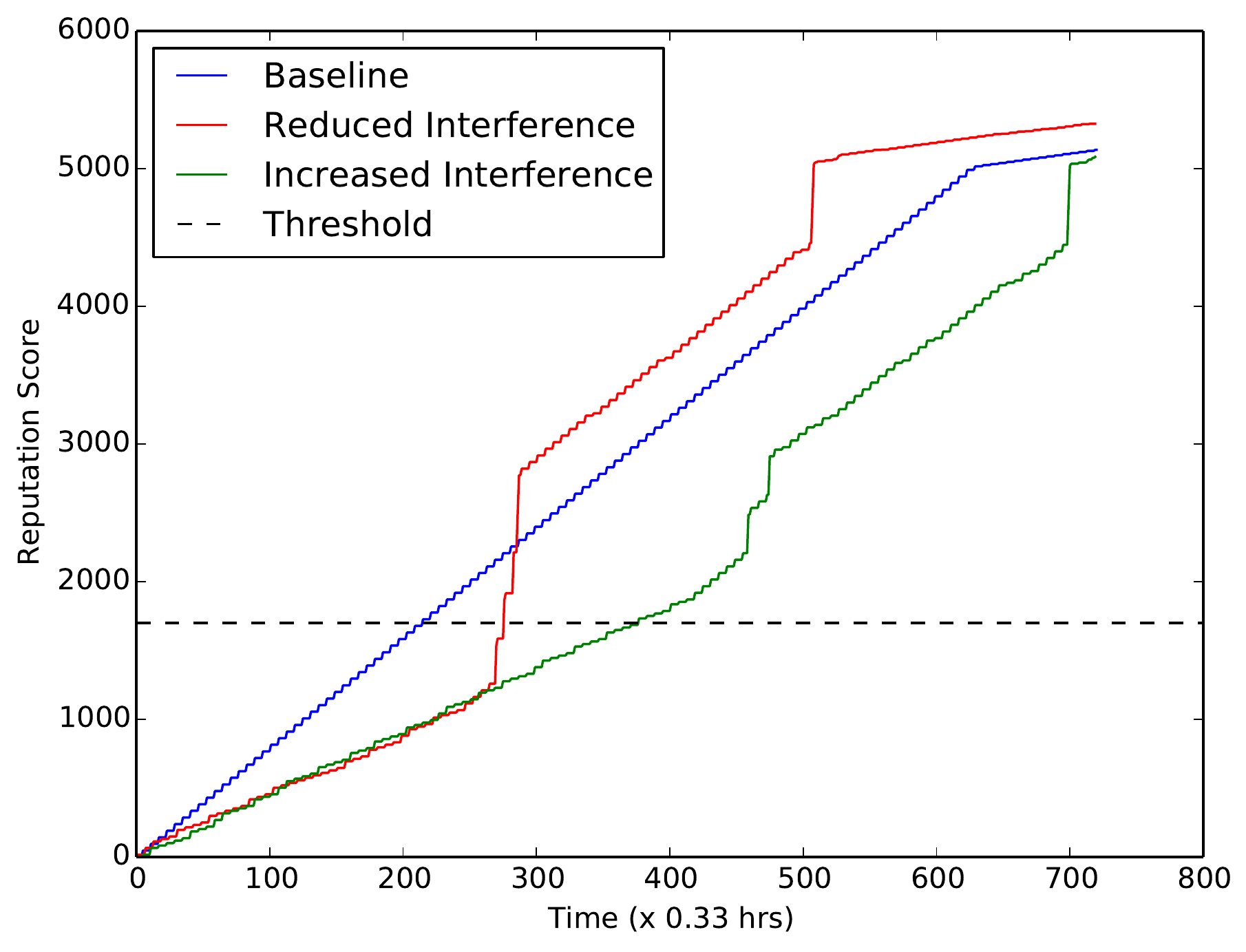}
    \end{center}
\vspace{-0.1in}
  \caption{\small{Reputation Score Growth}}
  \label{fig:RSG}
 \vspace{-0.1in}
\end{figure}

\noindent{\bf 1. Single Node Score Growth:} To provide visualization of the reputation score growth for a single node with the composite mobility model, we plot two cases: one in the absence and one in the presence of a severely active adversary. Using a scaled down version of the reputation manager (in terms of scoring), Figure \ref{fig:RSG} denotes the growth of the reputation for a node as a function of time, reaching a threshold of $1700$. The baseline case i.e. the absence of any adversarial interference or trusted locations converges in 70 hours. The second scenario where the node is in the presence of an adversary 43\% of the time i.e. the (increased) adversarial interference ($p$) is $0.43$. Thus, the node takes nearly double the time ($136$ hours) to reach the same threshold. In this extreme case where there is stunted growth of the reputation score, the protocol succeeds only because of score growth due to the trusted location distribution, where 1 of every 12 locations visited is trusted. 

\vspace{1mm}
\noindent{\bf 2. Number Of Exchanges:} Table \ref{table:avg} reports the average number of exchanges (NOE) required to reach a threshold of $3400$ for each of the mobility models across $10$ independent trials of the experiment and varied adversarial interference, where a subset (of size $100$) of participants are assigned the same priority. The Simple Traffic model has the largest NOE because participants move in a fixed direction, thereby reducing the chances of being in the vicinity of a trusted location frequently. On the other hand, the rate of mobility in the Downtown Manhattan model, while lower in some areas is higher in others. Thus, with increased mobility, the chances of being in the vicinity of a trusted location also increases, thereby lowering the NOE. 

\begin{figure*}
\centering
\subfigure{
\includegraphics[width=2.2in]{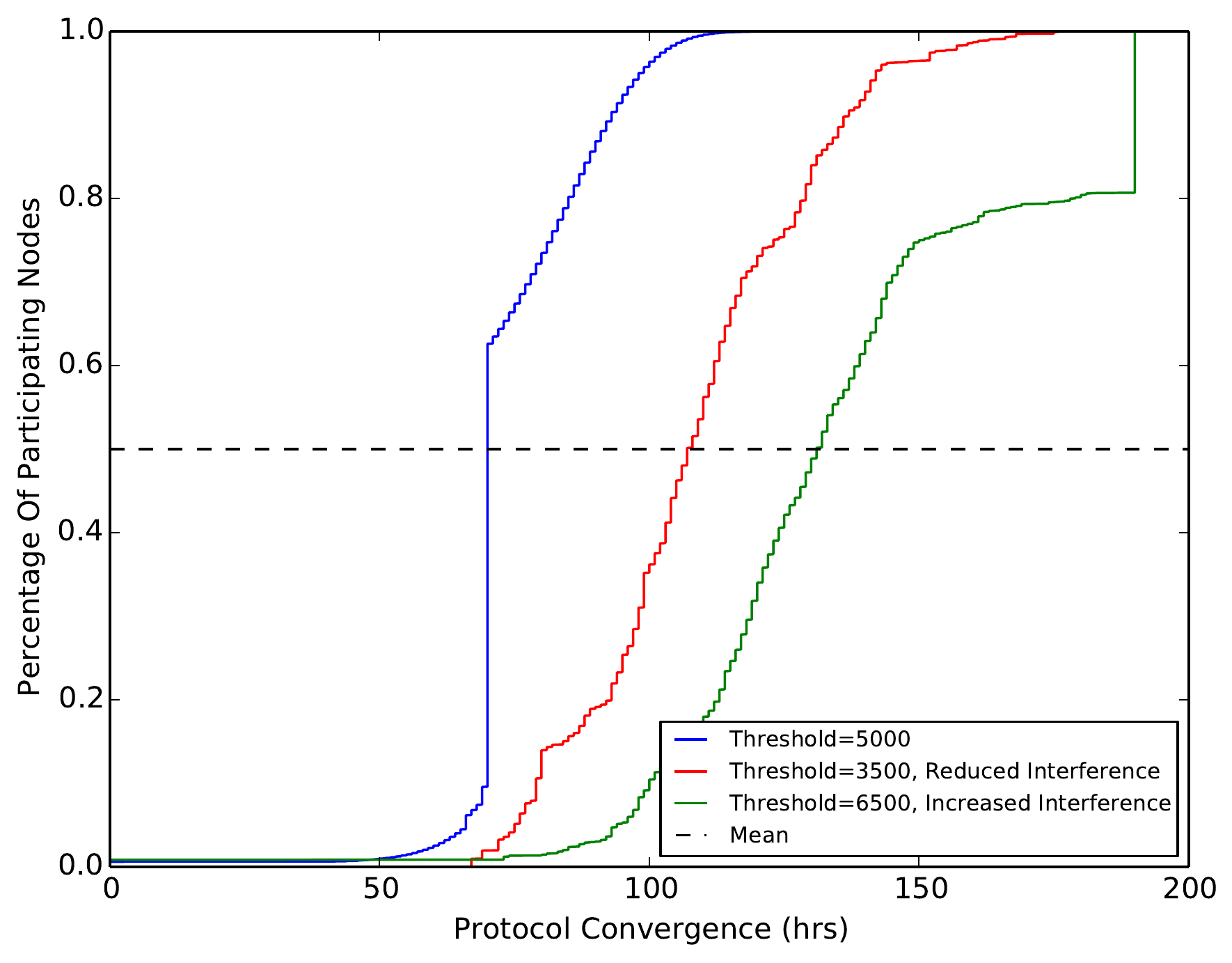}
\label{fig:cdf}
}
\subfigure{
\includegraphics[width=2.2in]{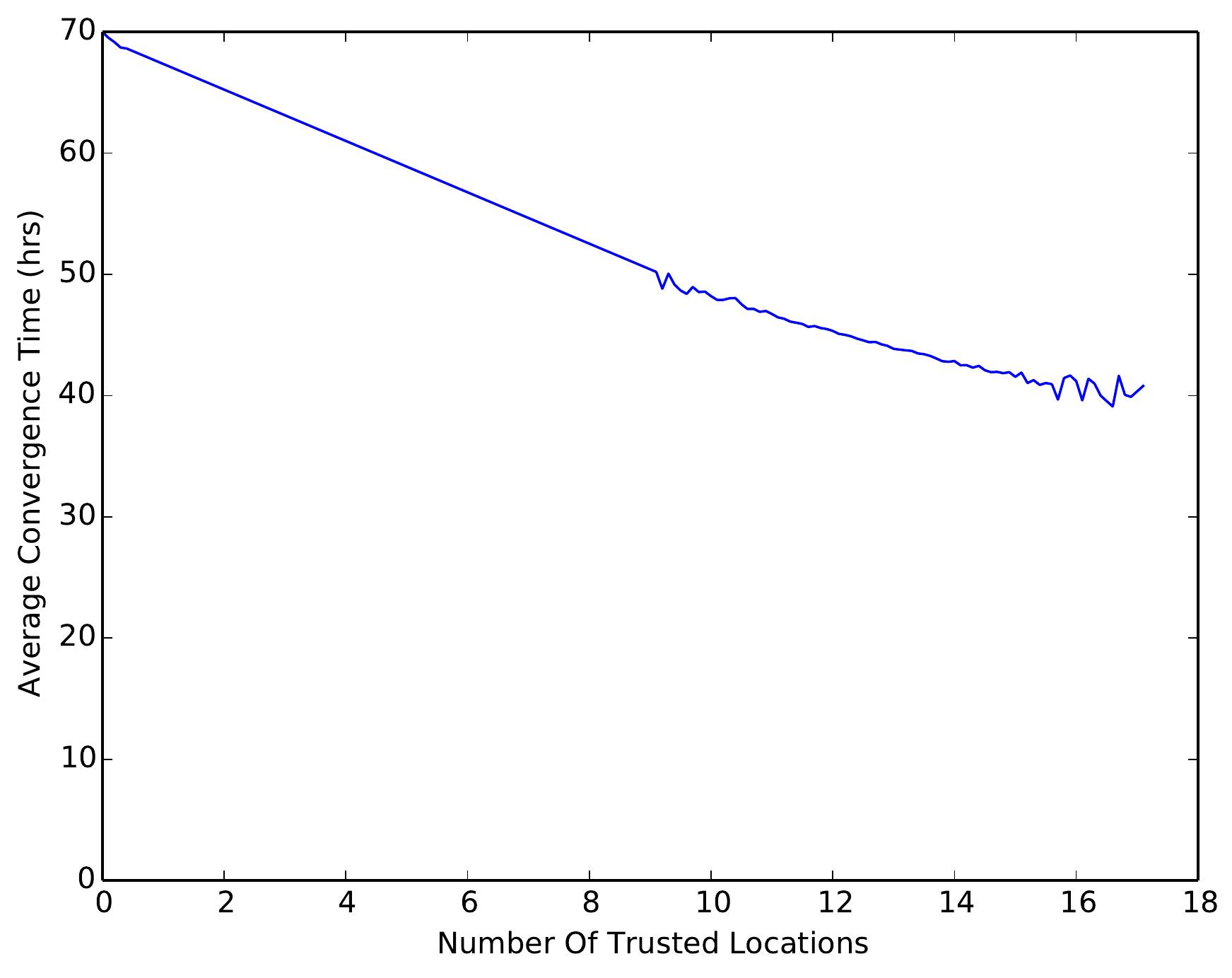}
\label{fig:trustedloc}
}
\subfigure{
\includegraphics[width=2.2in]{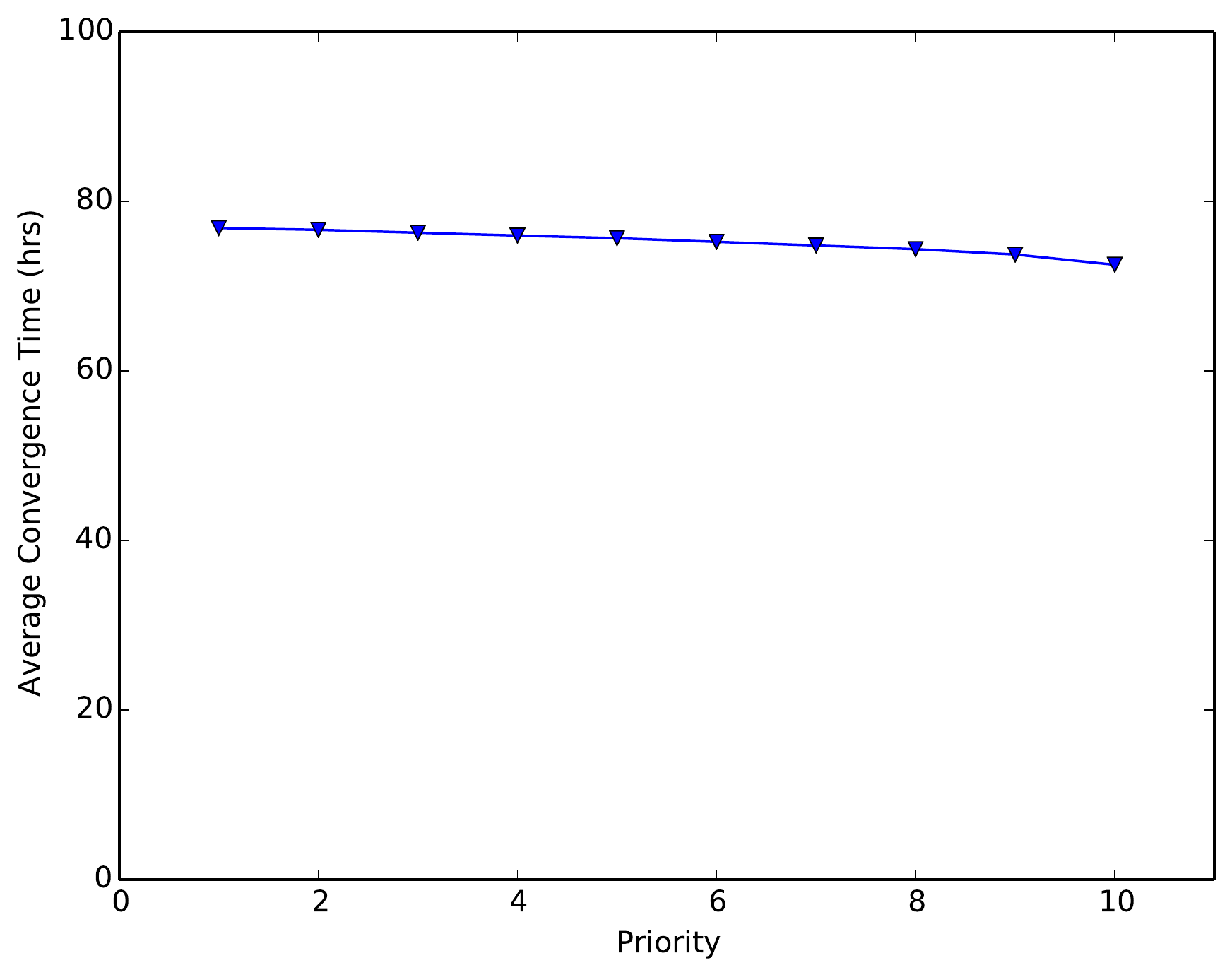}
\label{fig:PrivsTime}
}

\caption{Aggregate Statistics}
\end{figure*}

\vspace{1mm}
\noindent{\bf 3. Protocol Convergence:} We emulated key-exchanges at an hourly frequency for a period of 30 days, for all $10^4$ nodes in our experimental setup. For all these experiments, the threshold is set to $5000$. We observe that the time taken to successfully reach the predefined threshold varies based on: 
%(a) mobility (b) adversarial interference (c) number of trusted locations encountered (d) priority and batch of participating nodes. We explain each of them below, with empirical results:

\vspace{1mm}
{\em i. User Mobility:} Figure \ref{fig:cdf} suggests that in the baseline case, a very small fraction of total participating nodes (the bottom 0.0063 percentile) are immobile. We vary the reputation threshold and observe that the percentage of immobile users remains nearly constant across all experiments. The mean convergence time however, varies with the reputation threshold and amount of adversarial interference. For visualization purposes, we represent protocol failure using a convergence time of $190$ hours. We observe that nearly $20$\% of participating nodes fail with the severity of interference.

\begin{table}[h]
\centering
\begin{tabular}{| l | l | l | l |}
\hline 
\textbf{} & \small{\textbf{Interference}} & \small{\textbf{Convergence}} & \small{\textbf{$\lambda$}}\\ 
\hline
 Baseline & 0 & 70 hrs & 0\\ 
 \hline
 Reduced & 0.11 & 79.37 hrs & 0\\ 
 \hline
 Increased & 0.39 & 89.05 hrs & 25\%\\ 
 \hline
\end{tabular}
\caption {Interference \& Convergence}
\label{table:adv}
\end{table}

\begin{figure}[h]
  \begin{center}
      \includegraphics[width=.80\columnwidth]{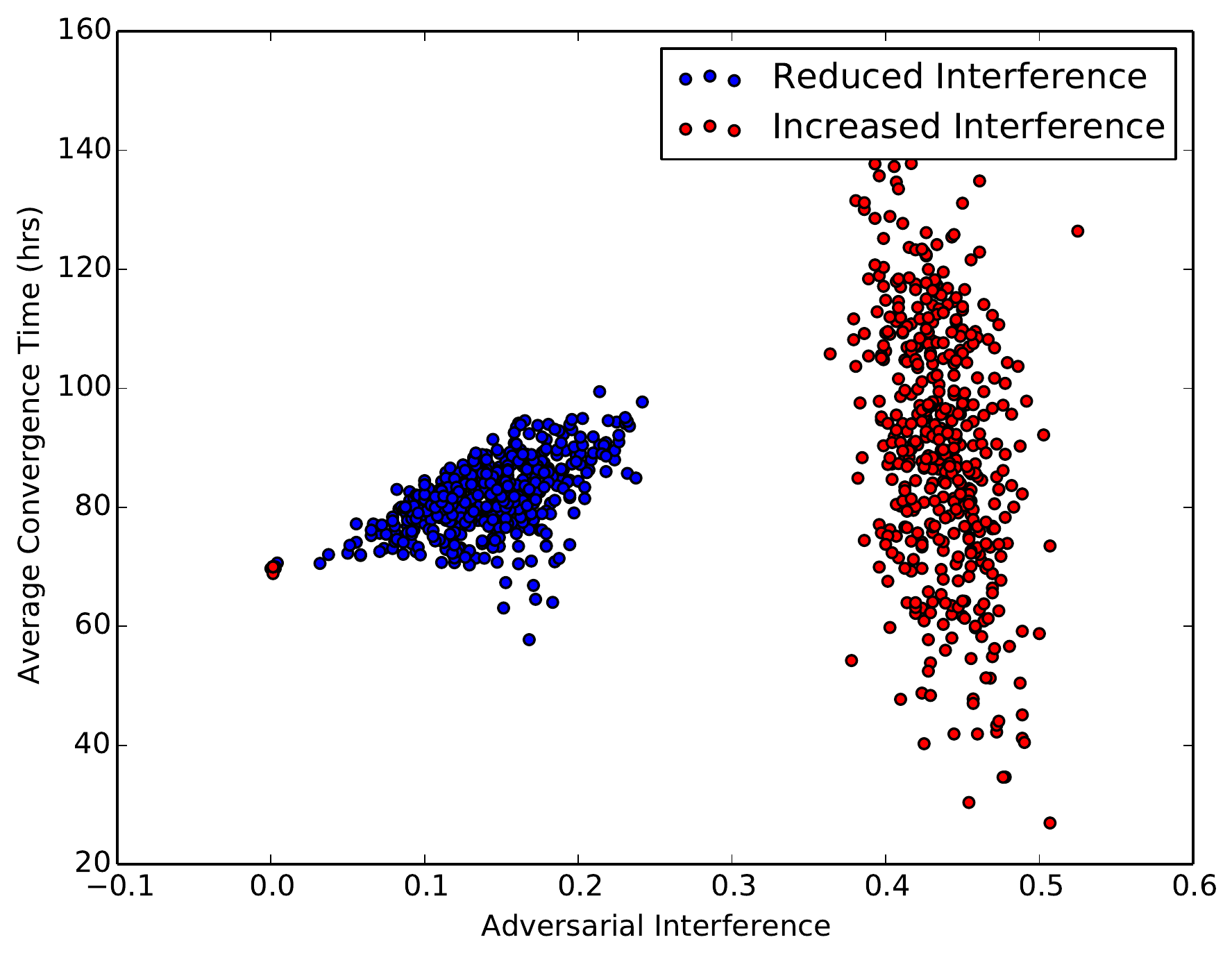}
    \end{center}
\vspace{-0.1in}
  \caption{\small{Convergence vs Adversarial Interference}}
  \label{fig:convadv}
 \vspace{-0.1in}
\end{figure}

\vspace{1mm}
{\em ii. Adversarial Interference:} We performed two sets of experiments, one with {\em reduced} adversarial interference with a maximum probability of 0.28 and another with {\em increased} adversarial interference of maximum probability 0.54. Note that even the maximum value of reduced interference is an overestimation of the actual value as an attack of such scale would require expansive economic resources. We sanction such (large) values to display the endurance of our protocol. We emulated $400$ trials of both experiments along with a baseline case for better comparison. As expected, increased interference affects a larger number of key-exchanges and thus increases the protocol convergence time, sometimes beyond $720$ hours. This increases the percentage of failure ($\lambda$) among participating nodes. On the other hand, reduced interference produces moderate delays towards convergence. Table \ref{table:adv} and Figure \ref{fig:convadv} reflect our claims.  

\vspace{1mm}
{\em iii. Number of Trusted Locations:} Across $400$ trials for a subset of users of size $10^2$ in the baseline case, we observed the impact of trusted locations towards protocol convergence. We achieved this by plotting the average convergence time against the number of trusted locations. Since the quick growth of reputation hinges on an increased number of such locations or cloud-based providers, we observe an expected direct correlation between the two parameters in Figure \ref{fig:trustedloc}. The small slope is due to the variation in trusted locations across zones; some have greater densities than others, and due to repeatedly visiting the same trusted location in a predefined window (of size 2 days in our experiments) disregarding the constraint for {\em unique trusted locations}.

\vspace{1mm}
{\em iv. Priority \& Batch of Participating Nodes:} The large scale emulation for $10^4$ nodes with reduced adversarial interference (p=0.28) was segregated and the convergence time was visualized as a function of the priority of the nodes. Since the reputation manager employs fixed priority scheduling, the protocol convergence time is inversely proportional to the priority of the participating nodes as in Figure \ref{fig:PrivsTime}. Outliers may exist in this plot because of two reasons, one being a cyclic-batch-wise score update module we have implemented as part of the reputation manager. This module ensures that no subset of nodes is starved for score updates due its reduced priority. The other reason is a module that reduces the maximum priority used as part of trusted-location based score increments. This reduction is done in a cyclic fashion in a periodic manner. However, these outliers have been normalized due to repeating this experiment $400$ times to account for sufficient randomness in batch, priority across nodes and adversarial interference, number of trusted locations across the emulation test-bed. By increasing the contribution due to trusted locations module, we can further increase the slope.

\begin{figure}[h]
  \begin{center}
      \includegraphics[width=.80\columnwidth]{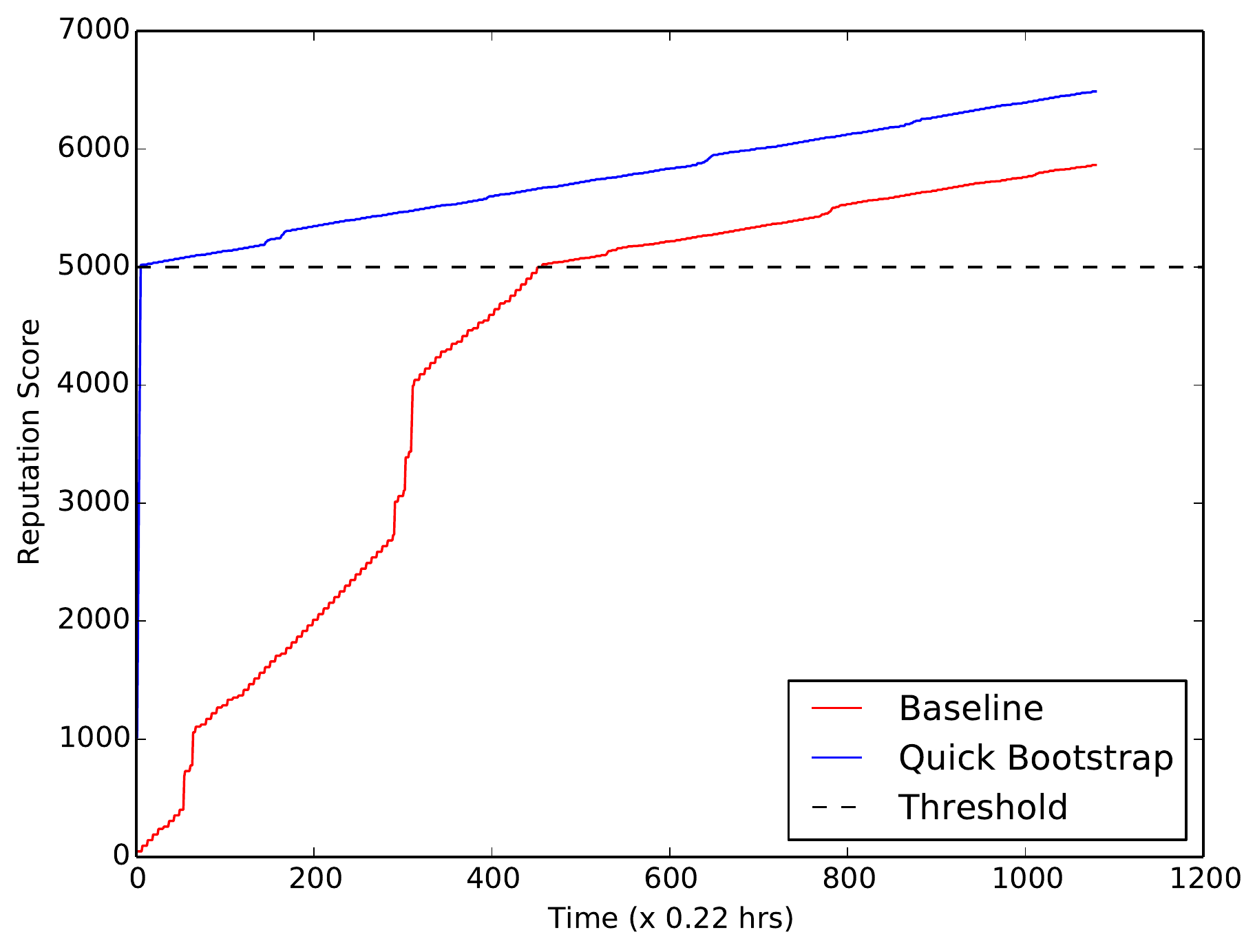}
    \end{center}
\vspace{-0.1in}
  \caption{\small{Quick Reputation Bootstrap}}
  \label{fig:qrb}
 \vspace{-0.1in}
\end{figure}

\vspace{1mm}
{\em v. Quick Reputation Bootstrap:} To emulate this phenomena (explained in \S \ref{tpiv}), we computed the reputation score with the trusted third-party and without (i.e. baseline case) but increased the key-exchange frequency to $40$ minutes. This frequency increase increases the baseline convergence because the average mobility (ratio of distance actually covered to the distance that can be covered) of nodes is lesser at a $0.67$ hour frequency when compared to an hourly frequency. From Figure \ref{fig:qrb}, we observe that the quick reputation bootstrap is faster by a factor of $80$, from $84$ hours to $1$ hour, highlighting its impact towards score growth. This decrease by nearly $1.9$ orders of magnitude is non-trivial, but comes with an implicit caveat. Staleness of the information at the third-party provider should be carefully measured.

%\subsection{Key Takeaways}
%
%\noindent{1.} Using Monte Carlo methods in our setup with $10^4$ nodes, only the bottom $0.0063$ percentile of users, on average, are immobile. 
%%A majority converge on time; only the top $20$ percentile experiencing abnormal delays.
%
%\noindent{2.} In an increased adversarial interference scenario, the protocol takes 27\% longer to converge. However, this extreme form of attack is highly unlikely. %In the reduced (yet overestimated) interference setup, the protocol always converges.
%
%\noindent{3.} As expected, the protocol converges faster with an increase in number of trusted locations and nodal priority. This is a consequence of our design decisions.
%% (\S \ref{setup}).
%
%\noindent{4.} The presence of quick reputation bootstrap reduces convergence time by nearly two orders of magnitude. This non-trivial improvement is very helpful in scenarios where users have limited time to establish trust.

\section{Applications}
\label{impl}

Our implementation of the SMI protocol is on top of the Android platform. The devices used in our implementation were Samsung Galaxy Young (GT-S5360) phones with Android v2.3.6. Standard APIs such as Bouncy Castle Java cryptography were used for cryptographic operations, with a key length of 2048 bits. Our implementation emulates physical synchronization with a trusted location using Bluetooth. 

\vspace{1mm}
\noindent{\bf Secure Messaging:} We have implemented a version of secure messaging on Android. On building sufficient reputation, the user can send signed messages to (a selected subset of) its contacts. This application has a SMS receiver component, which filters messages having a SMI header. Based on experiments conducted, we observe that the average time taken to send messages over the Etisalat network is $12.88$ seconds per message especially since encrypted messages will be split across several messages due to the signature length. The actual cryptographic operations consumed less than $0.06$ seconds or 0.4\% of the total time on a mobile device.
%, indicating the simplicity from the efficacy of standard cryptographic techniques used.

\vspace{1mm}
\noindent{\bf Secure Image Transfer:} We first generate a shared AES key and then encrypt the image using it, and transfer the encrypted image over the data channel. Given that a direct data path between mobile devices may not exist, we use an online cloud server as a synchronization point and exchange the data through it. The encrypted version of the shared key and {\em proof} of the file is shared over the SMS channel. This proof is a simple one-way hash of the byte-array of the image generated using SHA-1. On receiving the message, the recipient fetches the encrypted image, decrypts the AES key and then the image, and verifies the one-way hash proof. Based on experiments, we observe that the average time taken for the image transfer application is $27.65$ seconds out of which an average time of $20.42$ seconds was taken to upload the image. This extended period of time for upload is due to asymmetry in bandwidth provided.

\vspace{1mm}
\noindent{\bf Secure Transactions:} We developed an encrypted marketplace where mobile users share their keys using SMI. Then, using an online server as a synchronization point for a marketplace server, participants could place encrypted buy/sell orders. In essence, the central server is unaware of the exact transactions but enables a group of users to execute peer-to-peer transactions securely. The participants obtain three guarantees using SMI: (a) SMI establishes authenticity of the keys of the individual participants; (b) Every order placed in the marketplace was signed by the corresponding buyer or seller which enabled transaction-level trust between participants; (c) Since all transactions were encrypted, the central server could not learn much information about the individual transactions but could learn statistical patterns on the number of requests from each identity.

\section{Additional Considerations}
\label{extensions}

\noindent{\em Registering a new device:} Each cellular network can run their own key verification service where the device can generate its self-certifying key and initiate key-exchange with a {\em verifier} of the cellular network. This scheme of registering the user-generated public key on the cellular provider provides an expedited detection and mitigation strategy against network-level adversaries. In the extreme scenario where the adversary breaks the weak authentication, and spoofs the identity during first connect, it has to generate a different self-certifying key from the key generated by the device. Hence, the provider obtains two conflicting keys for the same identity and revokes access accordingly. 

\vspace{1mm}
\noindent{\em Key Revocation:} To provide better security with respect to the key itself, we ensure that the reputation of a user decays with time. A consequence of this is that participants of the protocol need to regularly participate in key-exchange to ensure liveness of the identity. On decay beneath a specific threshold, or on repeated blockage of interactions within an epoch, the previously generated (and used) self-certifying keys are revoked. A broadcast is sent to all associated users notifying them of this event. Thus, these ephemeral keys provide both forward and backward secrecy, as in the worst-case, a compromised key will provide no information about keys used prior, or to be used. 

\vspace{1mm}
\noindent{\em DDoS Alleviation:} SMI messages coupled with existing text message traffic can increase the network load on the cellular network. Enck et al. \cite{enck2005exploiting} suggest that to disturb cellular service, efficiently blanketing only a specific area with messages is sufficient so as to increase the probability of successful disturbance. Similar to the design of backoff protocols in wireless networks, SMI can use increasing delays in SMS messages and message losses as early signals to reduce the rate of messages from individual devices to reduce the network load. One additional strategy discussed earlier is to distribute the messages across the SMS and alternate channels to further reduce cellular network load.

%\vspace{2mm}
%\noindent{\em Global Network Adversary:} When faced with a computationally and economically unbounded adversary who can intercept a large fraction of cellular communications, key-exchanges can be severely hampered especially if the adversary can intercept all weak SIM authentication requests. The SMI protocol offers limited security guarantees in the face of such an all-powerful adversary. In the event where a global network adversary possesses limited access to intercepting communications within one cellular network, SMI users may be able to leverage device mobility across geographic locations to potentially discover the presence of conflicting key pairs for an identity in the event of a network level MitM attack from a global network adversary.

%\input{NSDI/7-Additional}

%\input{NSDI/9-Evaluation}
\section{Related work}

Despite the efforts of GSM to enhance their authentication security standards, many of the network security threats of cellular networks continue to remain. Toorani and Beheshti~\cite{toorani2008solutions} outline many of the weaknesses in the GSM authentication layer and propose simple fixes. The GSM specifications team~\cite{arkko2006extensible,haverinen2006extensible} proposed new cryptographic mechanisms to improve upon the early algorithmic vulnerabilities that plagued the original 2G authentication design. UMTS improved upon some of the security threats of the earlier GSM specifications to enhance integrity protection and authentication but even these extensions had several security problems~\cite{mobarhan2012evaluation}. The LTE authentication mechanism~\cite{bikos2013lte} builds upon the UMTS security model and introduces an key derivation hierarchy to enhance the security of the pre-shared keys; even these extensions suffered from several security problems~\cite{bikos2013lte, han2014security,tsay2012vulnerability}. Han and Choi~\cite{han2014security} demonstrate a threat against the LTE handover key management, involving a compromised base station. Tsay et al.~\cite{tsay2012vulnerability} find an attack on the UMTS and LTE AKA protocols using an automated protocol analyzer based on a computational model. Tang et al.~\cite{tang2013analysis} provide a detailed analysis of the security properties and vulnerabilities of different mobile authentication mechanisms.

Device pairing or key setup between two devices has been extensively studied
\cite{balfanz2002talking,castelluccia2005shake,bump,holmquist2001smart,lester2004you}. Specifically, relevant to SMI, we describe a few important secure encounter protocols. SDDR \cite{lentz2014sddr} provides secure encounters whilst enabling secure communication, providing selective linkability and silent revocation. SMILE \cite{smile} is a mobile “missed connections” application, that establishes trust between individuals who have shared a provable encounter. At the site of the encounter, MeetUp \cite{securesocial} proposes a visual authentication scheme using a trusted authority which attests a users public key to its picture. Groupthink \cite{groupthink} provides security primitives for users to count the number of participants and verify Short Authentic Strings (SAS). Secure Location Sharing (SLS) by Adams et al. \cite{clouds}, like SMI, uses multiple communication channels along with contextual question/answer protocols to prevent MitM attacks during device pairing. With the proliferation of WiFi gadgets and sensors that have susceptibility to different forms of MitM attacks, researchers have explored lower-layer signal abstractions to provide security over the wireless channel. Tamper-evident pairing (TEP) \cite{securepairing} is a protocol that provides simple, secure WiFi pairing and protects against MitM attacks without an out-of-band channel. SMI is only built for phones or for devices capable of sending a text message. SMI differs from these protocols in that it focuses establishing trust over the cellular channel building upon the weak authentication layer of GSM without heavy reliance on a PKI.

GAnGS \cite{gangs} exchanges the public keys of group members such that each member obtains the authentic public key of the other. However, continuous physical proximity is required. Moreover, the participant subgroup size can be constrained. SPATE \cite{spate} relies on visual channels and physical interactions. SPATE also requires a number of steps involving potentially inattentive end-users. SPATE was also designed for contact exchange in smaller groups (8 or fewer). Both GAnGS and SPATE enable bystanders to learn contact information, disclosing potentially sensitive private information. Our ephemeral key-pairs and counter-based encryption protect meta-data and ensure forward and backward secrecy. SafeSlinger \cite{safeslinger}, provides a system that leverages prior physical encounters to establish trust. The users communicate to determine a correct sequence that is common to all devices and select it to verify ephemeral keys, preventing users from simply proceeding. Incorporating mandatory verification may sacrifice asynchrony to ensure inattentive user resistance. Our system does not necessitate any pre-cursive encounter with other participants, and runs as a passive background process.

Keybase \cite{keybase} ensures safe retrieval of a particular public key using publicly readable social usernames. Each external service is used to post a signature proving that the account is bound to the named Keybase account. Onename \cite{onename} offers a verified "name" by registering a blockchain ID. Keybase and Onename can be viewed as independent third-party identity verifiers, which can co-exist with the SMI protocol.

\section{Discussion}
\label{ref:disc}

Independent of the results shown in \S \ref{eval}, we stress that the protocol functions in the absence of trusted locations/verifiers. The presence of these entities makes the protocol converge faster due their intrinsic trustworthy nature. This presents an interesting trade-off to be resolved by the end-user: {\em Can we relax our minimal trust assumptions for quicker usability?} The answer is also subject to the application that requires the trusted identity; applications requiring immediate trust require quicker convergence, while those that do not can permit a relaxed rate of growth. We believe that the time required for convergence is acceptable; organizations such as Google and Blockauth require similar timeframes to authenticate identities. We would also like to stress on the fact that most cellular providers in the USA (and other countries) provide unlimited text-messaging feature. This ensures that the number of messages required is not a bottleneck. This protocol is suited to work against a computationally and economically constrained network-level adversary with local scope; an adversary with omniscient network knowledge can repeatedly thwart any attempt to successfully build reputation and will hinder the progress of the protocol. Mitigating such an adversary is not within the scope of this work. For additional details not summarized in the paper, refer to the full technical report \cite{tech}.

%In the interest of space, details pertatining to selection of threshold, key revocation, DDoS alleviation and implementational benchmarks are absent and can be found in the extended version of the paper. 

\section{Conclusions}

Establishing secure identities in a decentralized manner is a challenging research proposition. In the Internet context, several decentralized solutions have been proposed and yet few have actually been adopted. The key takeaway from this paper is that SMI can augment weaker SIM security, building upon the very same weak authentication layer of cellular networks. 
%To summarize, the SMI protocol provides a repetitive spatio-temporally diverse key-exchange protocol that can quickly establish a high reputation for a self-certifying key generated by a mobile device. 
If SMI is broadly adopted, we believe it has very important ramifications for building end-to-end secure applications for mobile devices. We demonstrate these benefits using sample applications and evaluation of the security properties using standard mobility models. 

%As future work, we plan to build to use the SMI substrate to envision a more secure mobile Internet void of network-level security threats.

\newpage
\bibliographystyle{abbrv}
\bibliography{biblio,new}

\end{document}